\documentclass[a4paper,12pt]{revtex4}
\usepackage{amsmath}  % Para reconocer s\'{\i}mbolos AMS \'{o} AMS NOT
\usepackage{amssymb}  % Para reconocer s\'{\i}mbolos raros como \nRightarrow
\usepackage{dsfont}   % Para reconocer s\'{\i}mbolos de conjuntos (N,Z,R,C,I,...)

\begin{document}
\title{Do we finally understand Quantum Mechanics?}
\author{Alberto C. de la Torre}\email{delatorre@mdp.edu.ar}

\affiliation{Universidad Nacional de Mar del Plata\\Argentina}

\begin{abstract}
The ontology emerging from quantum field theory and the results
following from Bell's theorems allowed the development of an
intuitive picture of the microscopic world described by quantum
mechanics, that is, we can say that we understand this theory.
However there remain several aspects of it that are still mysterious
and require more work on the foundations of quantum mechanics.
\\ Published: arXiv:1605.00672 - %Rev. \textbf{77}, 22222 (2016).
\end{abstract}
\maketitle
\section{Introduction}
Fifty years ago R. Feynman said ``nobody understands quantum
mechanics''\cite{fey} characterizing the intellectual mood of that
time. Due to his deserved authority and the undeniable intelligence
of the big number of scientist that failed to develop a definite
interpretation of quantum mechanics, the pessimist idea that we
would \emph{never} understand it was established. Fortunately this
pessimism is perhaps unfounded and today we may have reached a level
of understanding sufficient for the development of an intuitive
picture of the physical systems described by quantum theory.

The essential developments that allowed this understanding are the
consequences of several ideas related to Bell's theorems and the
emergence of an ontology consistent with quantum field theory. In
this work we will present the main features of this progress towards
a final understanding of quantum mechanics. Much progress has been
done in the last decades but there are still remaining mysteries to
be understood therefore we can answer the question in the title by a
``yes but\ldots''

Although most of the material presented in this work is well known
for experts in quantum mechanics these issues are not usually
present in textbooks and lectures.  This work is therefore a useful
complement in the teaching of quantum theory because it brings
intuitive insights and also presents some unsolved questions as
possible research subjects in the foundations of quantum mechanics.

\section{Einstein Dilemma}
In a pioneering contribution that started an important field of
research in the foundations of quantum mechanics\cite{epr}  A.
Einstein, B. Podolsky and N. Rosen raised the question of the
completeness of quantum mechanics. An actualized reformulation of
this issue can be posed as the \emph{Einstein dilemma:} are the
quantum mechanical distributions of the values that an observable
can take in a particular state of a system of an ontological or a
gnoseological nature? Let us analyse these choices. For any
observable $A$ of a system, position, momentum, angular momentum,
etc., quantum mechanics provides a distribution with an expectation
value $\langle A\rangle$ and a width $\Delta_{ A}$ that
characterizes the uncertainty or indeterminacy of the observable in
the given state. We can take two options concerning the nature of
this distribution: \emph{gnoseological} or \emph{ontological}; that
is, are the uncertainties or indeterminacies in our knowledge of the
system or in the system itself?

We can think that each observable has some definite and exact value
in the system, called the \emph{putative value}\cite{isham}, that
quantum mechanics is unable to predict in general. In this case the
distribution represents our ignorance of the reality of the system:
it is a problem of our knowledge and therefore the distribution is
of a gnoseological nature. Since certainty is an attribute of
knowledge, we can call $\Delta_{ A}$ the \emph{uncertainty} of the
observable in the given state. If this is so, quantum mechanics is
not a complete theory and we immediately ask if there is a better
one, generically called \emph{hidden variable theories}, that may
predict these exact values assigned to the observables.

In the opposite interpretation, we assume that the observable does't
have a unique and sharp value and, instead, it is unprecise or
diffuse by nature: it is not a problem of our knowledge but of the
system itself with blurred observables. Correspondingly $\Delta_{
A}$ characterizes an \emph{indeterminacy} of the observable in the
given state.

\section{For whom the Bell tolls}
For the classically oriented intuition, the gnoseological option is
less traumatic. In fact, this was the choice of Einstein, Podolsky
and Rosen, and an intense research activity in hidden variables
began. However severe difficulties appeared with them. The first was
a ``no go'' theorem by von Neumann\cite{vneum} showing that such
theories could not exist, but this result was disproved by a
counterexample by Bell\cite{bell0}.

The next important result was a theorem by Bell\cite{bell0} and
another by Kochen and Specker\cite{koch} that proved that \emph{the
existence of non contextual putative values is in contradiction with
the formalism of quantum mechanics.}

In the proof of these theorems it is assumed that the putative
values do not depend on the context. This requires a detailed
explanation: in the description of a physical system we choose a set
of commuting observables in order to fix the state, that is, we
choose a context. For instance, the position of a particle along one
direction $X$ and the momentum along an orthogonal direction $P_{y}$
or $P_{z}$ (they commute with $X$) or total angular momentum $J^{2}$
and its projection along one arbitrary direction $J_{z}$, or the
position of one particle and the momentum of another particle. Any
observable can belong to different contexts and it is a very
reasonable assumption to think that the putative value assigned to
it does not depend on the context. After all, the context is decided
by the theoretical physicist in his office and this should have no
effect on the reality of the system, that is, the putative value are
also required to be context independent.

After the appearance of the original proofs by Bell and
Kochen-Specker several examples of the incompatibility of quantum
mechanics with the existence of putative values were presented
involving spin systems in different Hilbert space
dimension\cite{per, pen, mer, per1,cab} and also concerning position
and momentum observables\cite{dlt}. An elegant and simple proof of
high didactic value\cite{cass}, based on the geometrical structure
of the Hilbert space, was also produced.

We just saw that the \emph{formalism} of quantum mechanics forbids
the existence of definite values for the observables. A much more
important result is that \emph{the existence of non contextual
putative values is in contradiction with empirical reality.} This
 result follows from the experimental violation of Bell's
inequalities. Assuming the existence of definite numerical values,
even if they are unknown, Bell derived an inequality\cite{bell0}
concerning statistical averages of such values. Several equivalent
versions of this inequality, that could be tested experimentally,
were derived\cite{chsh} and finally the result of several
experiments\cite{exp} violated these predictions based on the
existence of non contextual putative values.

All these results related with Bell's contribution support then the
ontological interpretation of quantum mechanic distributions,
although the existence of \emph{context dependent} putative values
can not be excluded on logical ground, even though they are very
unlikely. In the particular case of the position observable, if we
accept that its distribution is ontological, then we must abandon
the image of a point particle (except when the particle is in an
eigenstate of the position operator, that is, with $\Delta_{x}=0$)
and think of it as an extended object: the ``particle'' is more
something like a ``field'' with all particle properties extended in
physical space. An electron in a hydrogen atom is not a point-like
particle located with some probability in a region called an
``orbital'' but is the orbital itself. However, in a spatial
interaction or in a measurement the electron collapses (see later)
and presents a ``point-like'' characteristic. Anyway, the
localization or the ``point-likeness'' has a limit because one can
show by a (dangerous) heuristic argument involving the uncertainty
position-momentum relation that the location indeterminacy can not
be smaller than the Compton length of the particle $\lambda
=\hbar/mc$.

The view of particles as extended objects is compatible with, and
suggested by, quantum field theory, perhaps the most successful
theory in physics.

\section{quantum field theory ontology }
Quantum field theory, conciliating quantum mechanics and special
relativity, was originally developed as a quantum electrodynamic
theory and was later generalized to all interactions (except
gravity) to achieve a successful description of the time evolution
and interaction of all the particles in the Standard Model: quarks,
leptons and intermediary bosons. There are many good books at the
advanced undergraduate and graduate level for his theory\cite{wein}.
However for the purpose of this work we don't need all mathematical
details, that sometimes blur the essential features of the theory,
and a ``minimal quantum field theory''\cite{dltQFTIQM} is
sufficient. The particle ``wave function'' of quantum mechanics
$\psi(x,t)$ becomes in quantum field theory the amplitude for the
creation of such a particle at the space-time point $(x,t)$. A
characteristic feature of this theory is the expansion of all
amplitudes in terms of creation and annihilation operators of all
possible particle properties.

These expansions allow, and suggest, an ontology where the quantum
field of a particle system is build by the creation, propagation and
annihilation of real entities ---virtual particles--- with ephemeral
existence because they don't satisfy the energy-momentum relations
of permanent particles $E^{2}-P^{2}=m^{2}$ (they are off the ``mass
shell''). These virtual particles are not only off the mass shell
but they also violate causality, because they can propagate in
space-like trajectories outside the light cone. One of the beauties
of quantum field theory is the restoration of global causality in
the quantum field that requires the existence of antiparticles.

The terms of the expansions can be represented by Feynman diagrams;
however, in this ontology, these are not only term in a perturbation
expansion but represent really occurring processes. Even the vacuum
becomes dynamical features with creation, propagation and
annihilation of all possible virtual particles and there are
empirical facts (Lamb shift and Casimir effect) supporting this
dynamical vacuum. The physical vacuum is different from ``nothing''.
An argument, of historical interest, that can be related with the
dynamic vacuum was produced by Johannes Kepler four centuries before
its discovery. Kepler argued that in vacuum there is nothing to
oppose the propagation of light and therefore its speed should be
unlimited. The argument is correct, but the premise is false: in
vacuum there is something ---a sea of virtual particles--- to oppose
the propagation of light.

The \emph{Quantum Field} of a particle system is then a physical
entity extended and evolving in space-time according to specific
equations of motion (Schr\"{o}dinger, Dirac, Klein-Gordon) made by an
infinite set of virtual particles.

\section{individuality loss }
One of the fundamental features of reality discovered by quantum
mechanics is the \emph{individuality loss}. In our perception of
macroscopic objects we take for granted that their individuality is
conserved: if we look at a stone, close our eyer for a second, and
observe it again, we never doubt that we are dealing with \emph{the
same} stone. This anthropocentric conviction can not be extrapolated
to the microscopic world. Identical \emph{classical} systems have an
individuality that is conserved through the time evolution and
interaction with other systems. So classical systems, even when they
are ``identical'', can be assigned an individual identity that is
conserved: they can have a name, an ID number, a licence plate.
Quantum mechanics requires a drastic conceptual change: \emph{the
individuality loss}. A set of five identical ``classical'' atoms is
countable (five in total) and numerable (the atom number one, the
number two,\ldots) but real atoms, necessarily described by quantum
mechanics, are countable but not numerable: if we artificially
assign a number to each atom, that is, if we assign an
individuality, we must correct this error by considering also all
possible permutations in the assignment. The individuality of a
particle is entangled with the individuality of all other identical
ones in the universe (although ``for all practical purposes'' a
cluster decomposition isolating a particular system from the rest is
possible to an extremely good approximation\cite{dlTM}).

An interesting metaphoric tool for a didactic presentation of the
identity entanglement in quantum mechanics is provided by some short
stories by Julio Cort\'{a}zar\cite{Cort} where the identity of some
characters are swaped.

In the ontology suggested by quantum field theory the individuality
loss is very natural because in this interpretation we are not
dealing with one, or two, or many particles as individual entities.
For instance, the quantum field for a one electron system, or for
several electrons system, is made up by the creation, propagation
and annihilation of virtual particles that are not assigned to any
of the individual electrons of the system: in a two electron field
there is no way to differentiate one electron from the other because
they are both simultaneously made by an active background of
ephemeral virtual particles with a mean value of \emph{two} for the
particle number observable, but each virtual component of the field
is not assigned to any one of the electrons.

\section{Decoherence in the classical limit}
Quantum mechanics is also applicable to macroscopic systems that do
not exhibit indeterminacies and other astonishing features of
microscopic quantum systems. Besides treating systems with
ontological indeterminacies, quantum theory can also describe
ensembles of systems with gnoseological uncertainties. The
appropriate tool is the \emph{statistical operator} $W$, also called
the \emph{density matrix}. Let us assume an observable $A$ with
eigenvalues $\{a\}$ and eigenvectors $\varphi_{a}$. A state $\psi$
where the observable has an ontological indeterminacy is given by
\begin{equation}\label{superpos}
   \psi = \sum_{a}f_{a}\varphi_{a} \ .
\end{equation}

Now we can think on an ensemble of systems where each member of the
set is in some state $\varphi_{a}$. If we have only a statistical
knowledge on how often these states are realized, that is, we know
the occupation probability $\lambda_{a}$ for each state, the state
of the ensemble is described by the statistical operator
\begin{equation}\label{mixt}
    W =  \sum_{a}\lambda_{a}P_{a} \ ,
\end{equation}
where $P_{a}$ is a projector in the state $\varphi_{a}$. Notice that
the \emph{Pure State} (\ref{superpos}) corresponds to an ontological
indeterminacy of the observable whereas the \emph{Mixed State}
(\ref{mixt}) implies a gnoseological uncertainty of the ensemble.
(Mixed states are also required to describe the state of a
\emph{subsystem} of a system in a known pure state.)

Macroscopic systems are almost never found in a superposition state
like (\ref{superpos}) because if they are forced in such a state, in
an extremely short time, estimated by $\frac{\hbar}{E}$, where $E$
is the total macroscopic energy of the system (a large value), there
is a transition from the superposition state $\psi$ to the mixed
state $W$. This transition is called \emph{decoherence}\cite{Decoh}
and it explains why quantum effects are not observed in macroscopic
systems.

This also solves some misunderstanding that appear when we ignore
decoherence and transfer a microscopic state to a macroscopic
system. The most famous example of this error is the Schr\"{o}dinger cat
argument: assume a quantum system in a superposition of two states,
spin up or down, or atom decayed or not, or particle at right or
left, etc. Assume also some amplifying mechanism that couples one of
these two states with the release of poison that kills a cat.
Clearly, to think that a real cat is in a superposition of
live-death is absurd and this will never appear in an experiment.
Due to decoherence, at all reasonable times in this cruel experiment
the cat will be alive \emph{or} dead and never alive \emph{and}
dead. The incorrect, but  popular, result that the cat is in some
live-death superposition is due to a misuse of quantum mechanics and
perhaps this motivated Stephen Hawking\cite{Hawk} to say ``When I
hear of Schr\"{o}dinger's cat, I reach for my gun'' paraphrasing the
opinion about ``culture'' attributed to several Nazi leaders but
with origin in a play of Hanns Johst\cite{Johst}.

Quantum mechanics is also applicable to macroscopic systems and
their state is the result of the decoherence of the superposition
states: the ontological indeterminacies characteristic of quantum
mechanics become gnoseological uncertainties in macroscopic systems.

\section{what remains mysterious}
The image that quantum field theory suggests for a particle in
space-time is quite intuitive and is also compatible with many
quantum features that where once considered counterintuitive. There
remain however several features of quantum mechanic that resist an
intuitive explanation. Some of them are the measurement problem, the
position-momentum relation, the quantization of rotations, and
several other. It is an open question whether unexpected future
developments will produce an explanation of these features or they
will just become familiar by getting used to them although no deep
understanding may never appear. After all, ``understanding'' is a
human mental state conditioned by our brain that has reached its
state after a few million years of evolution but it might not
necessarily be adequate for the microscopic world.

\subsection{measurement}
The essential difficulty in understanding the process of
measurement\cite{LonBau} is the mechanism by which the ontological
indeterminacy of an observable in a physical system becomes a
gnoseological uncertainty: the ``collapse''. This collapse is
acausal and indeterministic and we don't know if this is a fact of
nature, difficult to accept for our ``classical'' mind, that we must
just accept or if we will sometimes be able to explain it.

Let us assume a quantum system in a state $\psi$ given in terms of
the eigenvectors of an observable $A$ as in Eq.(\ref{superpos}).
Now, if we decide to measure $A$, we put the system in interaction
with a measurement apparatus that is necessarily a macroscopic
system. As it happens with Schr\"{o}dinger's cat, the hole system
decoheres: the quantum system goes to one of the eigenstates
$\varphi_{a}$ and the display of the apparatus shows the eigenvalue
$a$. It is impossible to predict which one of the states will result
(indeterminacy) and we can only give a probability for this,
$\lambda_{a}=|f_{a}|^{2}$, neither can we give a hamiltonian that
describes the time evolution during the collapse (acausality).

The essential difference between a measurement in a macroscopic
classical system and a corresponding one in a quantum system is that
in the classical case, the measurement informs us about a
\emph{preexistent} property of the system, whereas in a quantum
system, there is no preexistent value for the observable and the
measurement \emph{forces} the system in one of the eigenstates of
the observable.

\subsection{space-time and energy-momentum relation}
There are two fundamental perspectives in the consideration of
physical reality: the space-time and the energy-momentum view, that
is, kinematics and dynamics. Whereas space and time are immediately
related to our sense perception and are therefore intuitive,
energy-momentum require a definition and, in classical physics, are
related to the concept of \emph{matter in movement}. So we define
$E=\frac{1}{2}mv^{2}$ and $p=mv$ and their relativistic extension
$E^{2}-P^{2}=m^{2}$ and $P=\gamma mv$. These intuitive relations are
retained and are compatible with the mathematically more abstract
formulations where momentum is given as the Legendre transformation
in the transition from the lagrangian to the hamiltonian as well as
the Poisson bracket relation and the view of energy-momentum as
generators of space-time translations. The intuitive components are
retained in the powerful abstract mathematical formalism of
classical physics. In quantum mechanics position and momentum become
somehow incompatible, they ``don't commute''. The intuitive view as
matter in movement is lost, but the abstract commutation relations,
or equivalent, the generators of the group of translation are
retained. Energy-momentum are then related to space-time  by the
Fourier transformations and it would be important to recover an
intuitive explanation for this mathematical formalism. The proof
that Fourier transformation is compatible with a postulated
position-momentum (being-becoming) symmetry principle\cite{dlt9}
shreds some light to the problem but is not conclusive enough to
make it intuitive.

\subsection{quantization of energy and rotations}
The quantization of energy and rotations are an outstanding feature
of quantum mechanics that we have accepted by accustomation, we just
got used to it, but remain counterintuitive and are not really
understood. Of course, they are an unavoidable consequence of the
mathematical formalism of the theory and have their root in the
position-momentum commutation relations; however this does not give
an intuitive understanding.

A long hope, inspired by general relativity, is to find an
explanation of quantum mechanics based on a particular structure of
space-time. This ``pregeometry'' should explain the quantization of
rotations, and the resulting energy quantization, as self-consistent
possibility.

\subsection{state determination}
Position and momentum are the unique observables of a spinless
particle moving in space, all other observables are functions of
them. Therefore it came as a surprise, indicative of some missing
understanding of the reality of the system, that the complete
knowledge of the position distribution $|\psi(x)|^{2}$ and momentum
distribution $|\phi(p)|^{2}$ (where $\phi$ and $\psi$ are related by
Fourier transformation) \emph{is not} sufficient in order to
determine the state of the system. This question, initially raised
by Pauli\cite{pau}, triggered an intense investigation on the
necessary and sufficient information needed for an unambiguous state
determination, an unsolved problem in quantum
mechanics\cite{wei1,wei2,qinf,unbiaBas1,unbiaBas2,dlt4}. It has been
conjectured, but unproved, that the correlation observable $C=XP+PX$
might provide the missing information for complete state
determination\cite{dlt10}.

\subsection{compatibility with general relativity}
The greatest cultural debt of theoretical physics is \emph{quantum
gravity}, a theory enclosing quantum mechanics and general
relativity. There are several approaches to such a theory but they
have not reached sufficient development for an established theory.
Quantum mechanics and general relativity are incompatible at a very
fundamental level and therefore the new theory will bring profound,
perhaps revolutionary, concepts and ideas and certainly a different
understanding of quantum mechanics. The incompatibility of quantum
mechanics and general relativity arises from the fact that general
relativity requires a precise, sharp, definition of energy-momentum
at every precise space-time point, that is, the energy-momentum
tensor that determines the space-time structure through Einstein's
equations. However the uncertainty principle of quantum mechanics
forbids the simultaneous and precise definition of these quantities.

Quantum gravity is necessary for the description of physical systems
in unusual extreme conditions where quantum fluctuations become self
devouring black holes, like in the first $10^{-43}s$ (Planck time)
of the universe or the region $10^{-35}m$ (Planck length) around the
center of a black hole ``the undiscovered country from hose bourns
no traveller returns''\cite{Shakes}. Although the applicability of
such a theory is extremely small, the cultural gap is big and
physics will not rest until quantum gravity is found.

\section{conclusion}
 Progress made in the second half of the XX century have clarified
 several aspects of quantum theory and we can today imagine
 intuitively the microscopic world. So we can think of a hydrogen
 atom as an extended field for the electron ---the orbitals--- with
 virtual photons binding it to the proton, that in its turn is a
 field made of three quark fields in a sea of gluons and other intermediary bosons. All
 this in a fascinating and beautiful swarm of virtual particles.

 There are however many unsolved questions that make the research in
 the foundations in quantum mechanics a relevant scientific
 activity.


\begin{thebibliography}{99}

\bibitem{fey}
R. Feynman. {\em Character of Physical Law.} MIT Press (1967).

\bibitem{epr} A. Einstein, B. Podolsky, N. Rosen.
``Can quantum mechanical description of physical reality be
considered complete'', Phys. Rev. {\bf 47}, 777-780 (1935).

\bibitem{isham}
C. J. Isham. \emph{Lectures on Quantum Theory. Mathematical and
structural foundations}. Imperial College Press, London, (1995).

\bibitem{vneum}
J. von Neumann, \emph{Mathematische Grundlagen der Quantenmechanik,}
Springer, Berlin, (1932); English translation: \emph{Mathematical
foundations of quantum mechanics}, Princeton Univ. Press, (1955),
Chapter IV.1,2.

\bibitem{bell0}
J. Bell. ``On the problem of hidden variables in quantum theory'',
Rev. Mod. Phys. \textbf{38}, 447-52 (1966).

\bibitem{koch}S. Kochen and E. P. Specker. ``The problem of hidden
variables in quantum mechanics,''  J. Math. Mech. \textbf{17}, 59-88
(1967).

\bibitem{per}A. Peres, ``Two simple proofs of the Kochen-Specker theorem,''J. Phys. A
\textbf{24}, L175 (1991).

\bibitem{pen}R. Penrose, ``On Bell non-locality without probabilities: some curious
geometry,'' in \emph{``Quantum Reflections''} Pgs.1-27. J. Ellis and
D. Amati Eds. (Cambridge University Press, Cambridge, 2000).

\bibitem{mer}N. D. Mermin, ``Hidden variables and the two theorems of
John Bell,'' \emph{Rev. Mod. Phys.} \textbf{65}, 803-815
(1993).

\bibitem{per1}M. Kernaghan and A. Peres, ``Kochen-Specker
theorem for eight-dimensional space,'' \emph{Phys. Lett. A}
\textbf{198}, 1-5 (1995).

\bibitem{cab}A. Cabello, J. M. Estebaranz and G.
Garc\'{\i}a-Alcaine, ``Bell-Kochen-Specker theorem: a proof with 18
vectors'' \emph{Phys. Lett. A} \textbf{212}, 283-287 (1996).

\bibitem{dlt} A. C. de la Torre, ``Observables have no value:
a no-go theorem for position and momentum observables'' Found. of Phys. \textbf{37},
1243-1252 (2007)

\bibitem{cass} A. Cassinello and A.
Gallego, ``The quantum mechanical picture of the world,'' \emph{Am.
J. Phys.} \textbf{73}, 273-281 (2005).

\bibitem{chsh}
J. Clauser, M. Horne, A. Shimony, R. Holt,  ``Proposed Experiment to
Test Local Hidden-Variable Theories''  Phys. Rev. Lett. \textbf{23},
880–884  (1969).

\bibitem{exp} S. J. Freedman, J. F. Clauser,  ``Experimental test of local hidden variable
theories'', Phys Rev. Lett.  \textbf{28}, 938-941 (1972).\\
A. Aspect J. Dalibard and G. Roger. ``Experimental test of Bell's
inequalities using time varying analysers'', Phys. Rev. Lett.
\textbf{49}, 1804-1807 (1982).

\bibitem{wein} See for instance S. Weinberg.
\emph{The Quantum Theory of Fields}. Cambridge University Press
(1995).

\bibitem{dltQFTIQM}
A. C. de la Torre. ``The quantum field theory interpretation of
quantum mechanics'' arXiv: 1503.00675

\bibitem{dlTM}
A. C. de la Torre, H. O. M\'{a}rtin.  ``Distinguishing identical
particles and the correct counting of states'' Eur. J. Phys.
\textbf{30}, 467-475 (2009).

\bibitem{Cort} J. Cort\'{a}zar. ``Lejana'' in \emph{Bestiario} (1951); ``La noche boca
arriba'' and ``Axolotl'' in \emph{Final del juego} (1956)

\bibitem{Decoh}
E. Joos,  H. D.  Zeh. ``The emergence of classical properties
through interaction with the environment''  Z. Phys. B \textbf{59},
223-230 (1985).\\
W. H. Zurek. ``Decoherence and the transition from quantum to
classical'' Phys. Today \textbf{44}, 36 (1991).
\\
M. Schlosshauer. ``Decoherence, the measurement problem, and
interpretations of quantum mechanics'' Rev. Mod. Phys. \textbf{76},
1267-1305 (2005).

\bibitem{Hawk} Reported by Timothy Ferris in \emph{The Whole Shebang: A State-of-the-Universe(s)
Report}. Simon \& Schuster (1997).

\bibitem{Johst} H. Johst \emph{Schlageter} (1933).

\bibitem{LonBau}F. London, E. Bauer.
\emph{La th\'{e}orie de l'observation en m\'{e}canique quantique.} Hermann,
Paris (1939).

\bibitem{dlt9}
A. C. de la Torre. ``The position-momentum symmetry principle''
arXiv: 1311.2454

\bibitem{pau} W. Pauli, ``Die allgemeine Prinzipien de Wellenmechanik'',
Handb. Phys. \textbf{24} (2), 83-272 (1933).

\bibitem{wei1}
S. Weigert, ``Pauli problem for a spin of arbitrary length: A simple
method to determine its wave function'' Phys. Rev. A {\bf 45},
7688-7696 (1992).

\bibitem{wei2}
S. Weigert, ``How to determine a quantum state by measurements: The
Pauli problem for a particle with arbitrary potential'' Phys. Rev. A
{\bf 53}, 2078-2083 (1996).

\bibitem{qinf}M. Keyl,
``Fundamentals of quantum information theory'' Phys. Rep. A
\textbf{369}, 431-548 (2002).

\bibitem{unbiaBas1}  I. D. Ivanovic, ``Geometrical description of quantum
state determination'' J. Phys. A, 14, 3241-3245 (1981).

\bibitem{unbiaBas2}W.K. Wootters, and B.D. Fields,
``Optimal state-determination by mutually
unbiased measurements'' Ann. Phys. 191, 363-381 (1989).

\bibitem{dlt4}
D. M. Goyeneche and A. C. de la Torre, ``State determination: An
iterative algorithm'' Phys. Rev. A \textbf{77}, 042116 (2008).

\bibitem{dlt10}
A. C. de la Torre. ``On position, momentum and their correlation''
arXiv: 1308.4864

\bibitem{Shakes}
W. Shakespeare, \emph{Hamlet}, Act III, Scene 1

\end{thebibliography}
\end{document}